\documentclass[aps,prd]{revtex4}
\usepackage{eurosym}
\usepackage{amssymb}
\usepackage{amsfonts}
\usepackage{amsmath}
\usepackage{graphicx}

\begin{document}

\title{Quantum Tunneling of Massive Spin-1 Particles From Non-stationary
Metrics}
\author{I. Sakalli}
\email{izzet.sakalli@emu.edu.tr}
\author{A. \"{O}vg\"{u}n}
\email{ali.ovgun@emu.edu.tr}
\affiliation{Physics Department , Eastern Mediterranean University, Famagusta, Northern
Cyprus, Mersin 10, Turkey}
\date{\today }

\begin{abstract}
We focus on the HR of massive vector (spin-1) particles tunneling from
Schwarzschild BH expressed in the Kruskal-Szekeres (KS) and dynamic Lemaitre
(DL) coordinates. Using the Proca equation (PE) together with the
Hamilton-Jacobi (HJ) and the WKB methods, we show that the tunneling rate,
and its consequence Hawking temperature are well recovered by the quantum
tunneling of the massive vector particles.
\end{abstract}

\keywords{Hawking radiation, Regular coordinates, Proca equation , Massive
vector particles, Quantum Tunneling}
\pacs{04.62.+v, 04.70.Dy, 11.30.-j}
\maketitle

\section{Introduction}

One prediction of the theory of general relativity (GR) devised by Einstein
involves BHs. In principle, a BH is classically defined by an area of space
called the "event horizon", where everything is swallowed. Beyond the event
horizon, matter and light flow freely, but as soon as the horizon's
intangible boundary is crossed, matter and light become trapped. So, a BH is
an invisible object (i.e., black), at least classically.

Stephen Hawking's prediction that a BH might not be completely black is
unarguably one of the important consequences of the quantum mechanics, when
integrated with GR \cite{Hawking1,Hawking2,Hawking3}. In particular, Hawking
proved that a semiclassical BH possesses a characteristic temperature of a
thermally distributed radiation spectrum, which is the so-called Hawking
radiation (HR) \cite{Hawking1}. Today, in the literature there exists
several derivations of the HR, which are proposed to strengthen this
staggering theory (see, for example, \cite%
{Unruh,DR,pw,sri,sh1,s1,Clement,Mazhari,s3,Jiang,s5}). Among those methods,
the quantum tunneling method (QTM) of Angheben \cite{angh1} and Padmanabhan 
\cite{sri,pad2} (with their collaborators) has garnered much attention (see 
\cite{Vanzo1} and references therein). QTM employes the complex path
integral analysis of Kerner and Mann \cite{mann1,mann2} in the HJ formalism,
which takes account of the WKB approximation \cite{wkb}. According to the
QTM, a wave propagator that is proportional to $\exp \left( \frac{i}{\hbar }%
S_{0}+S_{1}+O(\hbar )\right) $ is applied to the wave equation of the
tunneling particle under question. Here, each $S$ denotes the classical
action of the trajectory of the particles coming out/in from the horizon.

In particle physics, a vector boson is a boson with the spin-1. In
particular, the massive vector bosons \cite{RevPar} i.e., $W^{\pm }$ and $Z$%
\ particles (force carriers of the weak interaction) play a prominent role
in the confirmed Higgs Boson \cite{Atlas}. Nowadays, the detection of a
massive photon, which is the so-called \textit{Darklight} \cite{dark1,dark2}
has become very popular in the experimental physics since it is envisaged to
explain the dark matter \cite{dark3}. Furthermore, in theoretical physics,
HR of the massive vector particles in stationary BHs have also attracted
much attention (see, for example, \cite%
{kr1,kg1,kr2,zhong,a5,a2,ch,ch2,ga1,kon1,kon2,rosa,hu1}). However, the
number of studies regarding the HR of the spin-1 particles from the
non-stationary \textit{regular} metrics is very limited \cite{ijmpa}, and
hence those regular spacetimes deserve more research. Such an extension is
one of the goals of the present paper. For this purpose, we consider the PE 
\cite{kr1,proca,proca2} in the KS \cite{ks,ks2} and DL \cite{dl}
coordinates. Next we apply the QTM to the PEs, and obtain a set of
differential equations for each coordinate system. Those equations enable us
to get a coefficient matrix. After setting the determinant of the
coefficient matrix to zero, we get the action $S_{0}$, which is the leading
order in $\hbar $. Then, we show how one can compute the tunneling rate of
the vector particles in the non-stationary metrics, and recover the standard
Hawking temperature of the Schwarzschild BH.

The paper is organized as follows: In Sec. 2, we first give a brief
introduction about the Schwarzschild spacetime in KS coordinates. Then, a
detailed calculation of quantum tunneling of spin-1 particles near the KS
horizon is provided. Section 3 is devoted to the computation of the HR of
the Schwarzschild BH from the tunneling of the massive vector particles in
the DL coordinates. The PE with minimum length effect and conclusions are
presented in Sec. 4.

\section{Quantum Tunneling of Massive Vector Particles From Schwarzschild BH
in KS Coordinates}

The well-known Schwarzschild solution is in general described by the
coordinates $t$ and $r$ as follows 
\begin{equation}
ds^{2}=g_{tt}dt^{2}+g_{rr}dr^{2}+g_{\theta\theta}d\theta^{2}+g_{\varphi
\varphi}d\varphi^{2},  \label{1}
\end{equation}
where $g_{tt}=(1-2M/r)$, $g_{rr}=-(1-2M/r)^{-1}$, $g_{\theta\theta}=-r^{2}$,
and $g_{\varphi\varphi}=-r^{2}\sin^{2}\theta$. Herein, there is an event
horizon at $r=2M$, so that $g_{rr}$ blows up. On the other side when $r<2M$,
the $g_{tt}$ and $g_{rr}$ exchange their signatures, however the signatures
of\emph{\ }$g_{\theta\theta}$ and $g_{\varphi\varphi}$ are not affected.%
\emph{\ }Hence, $r$ becomes \textquotedblleft timelike\textquotedblright\
and $t$ becomes \textquotedblleft spacelike\textquotedblright\ inside the
event horizon. One could clear up this\ \textquotedblleft coordinate
singularity\textquotedblright\ problem by introducing the KS coordinates 
\cite{ks,ks2}:

\begin{equation}
ds^{2}=A(-d\tau^{2}+dR^{2})+r^{2}(d\theta^{2}+B^{2}d\varphi^{2}),  \label{2}
\end{equation}

where $B=sin(\theta)$ and the metric function $A$ is given by 
\begin{equation}
A=\frac{32M^{3}}{r}e^{-\frac{r}{2M}}.  \label{3}
\end{equation}

Metric (\ref{2}) covers the entire spacetime manifold of the maximally
extended Schwarzschild solution, and it is well-behaved everywhere outside
the physical singularity ($r=0$). The event horizon in the KS coordinates
corresponds to $\tau$ $=$ $\pm$ $R$, and the curvature singularity is
located at $\tau^{2}-R^{2}=1$. Furthermore, in this coordinate system the
Killing vector becomes

\begin{equation}
\xi^{\mu}=\left[ \frac{R}{4M},\frac{\tau}{4M},0,0\right] .  \label{4}
\end{equation}

The particle energy of a test particle is given by (in terms of the action $%
S $) \cite{doran,ding} 
\begin{equation}
E=-\xi^{\mu}\partial_{\mu}S=-\left( \frac{R}{4M}\partial_{\tau}+\frac{\tau }{%
4M}\partial_{R}\right) S.  \label{5}
\end{equation}

On the other hand, for a curved spacetime, the PE is governed by \cite{kr1}

\begin{equation}
\frac{1}{\sqrt{-g}}\partial_{\mu}(\sqrt{-g}\Psi^{\nu\mu})+\frac{m^{2}}{%
\hbar^{2}}\Psi^{\nu}=0,  \label{6}
\end{equation}

where $\Psi _{\nu }=(\Psi _{0,}\Psi _{1,}\Psi _{2,}\Psi _{3})$ and $m$
represent the spinor fields \cite{kr1,kg1,ijmpa} and mass of the spin-1
particle, respectively, and

\begin{equation}
\Psi _{\nu \mu }=\partial _{\nu }\Psi _{\mu }-\partial _{\mu }\Psi _{\nu }.
\label{7}
\end{equation}

Using metric (\ref{1}) in Eq. (\ref{6}) , we obtain the following set of
differential equations:

\begin{multline}
-\left( \hbar ArB\right)
^{-2}[m^{2}r^{2}AB^{2}\Psi_{0}+\hbar^{2}r^{2}B^{2}(-\partial_{RR}\Psi_{0}+%
\partial_{\tau
R}\Psi_{1})-\hbar^{2}B^{2}A(\partial_{\theta\theta}\Psi_{0}-\partial_{\theta%
\tau}\Psi _{2})  \label{8n} \\
+\hbar^{2}BA\left( \partial_{\theta}B\right) (\partial_{\tau}\Psi
_{2}-\partial_{\theta}\Psi_{0})-\hbar^{2}A(\partial_{\varphi\varphi}\Psi
_{0}-\partial_{\varphi\tau}\Psi_{3})]=0,
\end{multline}

\begin{multline}
\left( \hbar ArB\right)
^{-2}[m^{2}Ar^{2}B^{2}\Psi_{1}-\hbar^{2}r^{2}B^{2}(\partial_{\tau
R}\Psi_{0}-\partial_{\tau\tau}\Psi_{1})-\hbar^{2}B^{2}A(\partial_{\theta%
\theta}\Psi_{1}-\partial_{\theta R}\Psi_{2})  \label{9n} \\
+\hbar^{2}BA\left( \partial_{\theta}B\right)
(\partial_{R}\Psi_{2}-\partial_{\theta}\Psi_{1})-\hbar^{2}A(\partial_{%
\varphi\varphi}\Psi _{1}-\partial_{\varphi R}\Psi_{3})]=0,
\end{multline}

\begin{multline}
\left( \hbar\sqrt{A}r^{2}B\right) ^{-2}[m^{2}Ar^{2}B^{2}\Psi_{2}-\hbar
^{2}r^{2}B^{2}(\partial_{\theta\tau}\Psi_{0}-\partial_{\tau\tau}\Psi
_{2})-\hbar^{2}r^{2}B^{2}(\partial_{\theta R}\Psi_{1}-\partial_{RR}\Psi _{2})
\label{10n} \\
-\hbar^{2}A(\partial_{\varphi\varphi}\Psi_{2}-\partial_{\theta\varphi}\Psi
_{3})]=0,
\end{multline}

\begin{multline}
\left( \hbar\sqrt{A}r^{2}B^{\frac{3}{2}}\right) ^{-2}[m^{2}Ar^{2}B\Psi
_{3}-\hbar^{2}r^{2}B(\partial_{\varphi\tau}\Psi_{0}-\partial_{\tau\tau}%
\Psi_{3})+\hbar^{2}r^{2}B(\partial_{\varphi R}\Psi_{1}-\partial_{RR}\Psi
_{3})  \label{11n} \\
+\hbar^{2}AB(\partial_{\theta\varphi}\Psi_{2}-\partial_{\theta\theta}\Psi
_{3})+\hbar^{2}A\left( \partial_{\theta}B\right) (\partial_{\theta}\Psi
_{3}-\partial_{\varphi}\Psi_{2})]=0.
\end{multline}

Applying the WKB approximation \cite{ijmpa}: 
\begin{equation}
\Psi_{\nu}=c_{\nu}\exp\left[ \frac{i}{\hbar}S_{0}(\tau,R,\theta
,\varphi)+S_{1}(\tau,R,\theta,\varphi)+O(\hbar)\right] ,  \label{12}
\end{equation}
and taking the lowest order of $\hbar$, Eqs. (8-11) become

\begin{multline}
\left( -r^{2}B^{2}\left( \partial_{R}S_{0}\right) ^{2}-A\left( \partial
_{\varphi}S_{0}\right) ^{2}-m^{2}Ar^{2}B^{2}-AB^{2}\left( \partial_{\theta
}S_{0}\right) ^{2}\right) c_{0} \\
+ c_{3}A\left( \partial_{\tau\varphi}S_{0}\right) +c_{1}r^{2}B^{2}\left(
\partial_{R\tau}S_{0}\right) +c_{2}AB^{2}\left(
\partial_{\tau\theta}S_{0}\right) =0,  \label{13n}
\end{multline}

\begin{multline}
r^{2}B^{2}\left( \partial_{R\tau}S_{0}\right) c_{0}+c_{1}\left(
-r^{2}B^{2}\left( \partial_{\tau}S_{0}\right) ^{2}+A\left( \partial_{\varphi
}S_{0}\right) ^{2}+m^{2}Ar^{2}B^{2}+AB^{2}\left(
\partial_{\theta}S_{0}\right) ^{2}\right) -c_{3}A\left(
\partial_{R\varphi}S_{0}\right)  \label{14n} \\
-c_{2}AB^{2}\left( \partial_{R\theta}S_{0}\right) =0,
\end{multline}

\begin{multline}
r^{2}B^{4}\left[ \left( \partial_{\tau\theta}S_{0}\right) c_{0}-c_{1}\left(
\partial_{R\theta}S_{0}\right) \right] +c_{2}\{AB^{2}\left(
\partial_{\varphi}S_{0}\right) ^{2}  \label{15n} \\
-B^{4}r^{2}\left[ \left( \partial_{\tau}S_{0}\right) ^{2}-\left(
\partial_{R}S_{0}\right) ^{2}-m^{2}A\right] \}-AB^{2}c_{3}\left(
\partial_{\varphi\theta}S_{0}\right) =0,
\end{multline}

\begin{multline}
r^{2}B\left[ \left( \partial_{\tau\varphi}S_{0}\right) c_{0}-c_{1}\left(
\partial_{R\varphi}S_{0}\right) \right] -c_{2}AB\left( \partial
_{\varphi\theta}S_{0}\right)  \label{16n} \\
+Br^{2}\left[ m^{2}A-\left( \partial_{\tau}S_{0}\right) ^{2}+\left(
\partial_{R}S_{0}\right) ^{2}-\frac{A}{r^{2}}\left(
\partial_{\theta}S_{0}\right) ^{2}\right] c_{3}=0.
\end{multline}

Now, one can obtain a matrix equation $%
%TCIMACRO{\U{2124} }%
%BeginExpansion
\mathbb{Z}
%EndExpansion
\left( c_{0},c_{1},c_{2},c_{3}\right) ^{T}=0$ \cite{kr1,kg1} (the
superscript $T$ means the transition to the transposed vector, and $%
%TCIMACRO{\U{2124} }%
%BeginExpansion
\mathbb{Z}
%EndExpansion
$ represents a $4\times 4$ matrix) with the following non-zero elements:

\begin{align}
%TCIMACRO{\U{2124} }%
%BeginExpansion
\mathbb{Z}
%EndExpansion
_{11}& =\left[ -r^{2}B^{2}\left( \partial _{R}S_{0}\right) ^{2}-A\left(
\partial _{\varphi }S_{0}\right) ^{2}-m^{2}Ar^{2}B^{2}-AB^{2}\left( \partial
_{\theta }S_{0}\right) ^{2}\right] ,\   \notag \\
%TCIMACRO{\U{2124} }%
%BeginExpansion
\mathbb{Z}
%EndExpansion
_{12}& =%
%TCIMACRO{\U{2124} }%
%BeginExpansion
\mathbb{Z}
%EndExpansion
_{21}=r^{2}B^{2}\left( \partial _{R\tau }S_{0}\right) ,  \notag \\
%TCIMACRO{\U{2124} }%
%BeginExpansion
\mathbb{Z}
%EndExpansion
_{13}& =AB^{2}\left( \partial _{\tau \theta }S_{0}\right) ,\text{ \ \ }%
%TCIMACRO{\U{2124} }%
%BeginExpansion
\mathbb{Z}
%EndExpansion
_{31}=r^{2}B^{4}\left( \partial _{\tau \theta }S_{0}\right) ,  \notag \\
%TCIMACRO{\U{2124} }%
%BeginExpansion
\mathbb{Z}
%EndExpansion
_{14}& =A\left( \partial _{\tau \varphi }S_{0}\right) ,\text{ \ }%
%TCIMACRO{\U{2124} }%
%BeginExpansion
\mathbb{Z}
%EndExpansion
_{41}=r^{2}B\left( \partial _{\tau \varphi }S_{0}\right) ,  \notag \\
%TCIMACRO{\U{2124} }%
%BeginExpansion
\mathbb{Z}
%EndExpansion
_{22}& =\left[ -r^{2}B^{2}\left( \partial _{\tau }S_{0}\right) ^{2}+A\left(
\partial _{\varphi }S_{0}\right) ^{2}+m^{2}Ar^{2}B^{2}+AB^{2}\left( \partial
_{\theta }S_{0}\right) ^{2}\right] ,  \notag \\
%TCIMACRO{\U{2124} }%
%BeginExpansion
\mathbb{Z}
%EndExpansion
_{23}& =-AB^{2}\left( \partial _{R\theta }S_{0}\right) ,\text{ \ }%
%TCIMACRO{\U{2124} }%
%BeginExpansion
\mathbb{Z}
%EndExpansion
_{32}=-r^{2}B^{4}\left( \partial _{R\theta }S_{0}\right) ,  \notag \\
%TCIMACRO{\U{2124} }%
%BeginExpansion
\mathbb{Z}
%EndExpansion
_{24}& =-A\left( \partial _{R\varphi }S_{0}\right) ,\text{ \ }%
%TCIMACRO{\U{2124} }%
%BeginExpansion
\mathbb{Z}
%EndExpansion
_{42}=-\ r^{2}B\left( \partial _{R\varphi }S_{0}\right) ,  \notag \\
%TCIMACRO{\U{2124} }%
%BeginExpansion
\mathbb{Z}
%EndExpansion
_{33}& =AB^{2}\left( \partial _{\varphi }S_{0}\right) ^{2}-B^{4}r^{2}\left[
\left( \partial _{\tau }S_{0}\right) ^{2}-\left( \partial _{R}S_{0}\right)
^{2}-m^{2}A\right] ,  \notag \\
%TCIMACRO{\U{2124} }%
%BeginExpansion
\mathbb{Z}
%EndExpansion
_{34}& =-AB^{2}\left( \partial _{\varphi \theta }S_{0}\right) ,\text{ \ }%
%TCIMACRO{\U{2124} }%
%BeginExpansion
\mathbb{Z}
%EndExpansion
_{43}=-AB\left( \partial _{\varphi \theta }S_{0}\right) ,  \notag \\
%TCIMACRO{\U{2124} }%
%BeginExpansion
\mathbb{Z}
%EndExpansion
_{44}& =Br^{2}\left[ m^{2}A-\left( \partial _{\tau }S_{0}\right) ^{2}+\left(
\partial _{R}S_{0}\right) ^{2}-\frac{A}{r^{2}}\left( \partial _{\theta
}S_{0}\right) ^{2}\right] .  \label{17}
\end{align}%
Let us consider the following HJ solution for the action: 
\begin{equation}
S_{0}=Q(\tau ,R)+k\left( \theta \right) +j\varphi ,  \label{18}
\end{equation}

where $j$ denotes the angular momentum of the massive vector particle. Thus,
the determinant of $%
%TCIMACRO{\U{2124} }%
%BeginExpansion
\mathbb{Z}
%EndExpansion
$-matrix yields

\begin{equation}
\det 
%TCIMACRO{\U{2124} }%
%BeginExpansion
\mathbb{Z}
%EndExpansion
=-m^{2}r^{2}AB^{3}\{A\left[ B^{2}\left( \partial _{\theta }k\right)
^{2}+j^{2}\right] +r^{2}B^{2}\left[ m^{2}A-\left( \partial _{\tau }Q\right)
^{2}+(\partial _{R}Q)^{2}\right] \}^{3}.  \label{19n}
\end{equation}

The nontrivial solution for $\partial _{R}Q$ is obtained by the condition of
"$\det 
%TCIMACRO{\U{2124} }%
%BeginExpansion
\mathbb{Z}
%EndExpansion
=0"$ \cite{kr1}. Hence, after substituting $\partial _{\tau }Q=-\frac{4ME}{R}%
-\frac{\tau }{R}\partial _{R}Q$ [recall Eq. (\ref{5})] into Eq. (19), we
obtain

\begin{equation}
\partial _{R}Q_{\pm }=\frac{4EM\tau rB\pm R\sqrt{%
16E^{2}M^{2}r^{2}B^{2}-A(R^{2}-\tau ^{2})\left\{ \left[ \left( \partial
_{\theta }k\right) ^{2}+m^{2}r^{2}\right] B^{2}+j^{2}\right\} }}{(R^{2}-\tau
^{2})Br}.  \label{20}
\end{equation}

where $+$ $(-)$\ corresponds to the outgoing (incoming) massive vector
particles. The definite integration of $Q$ is given by 
\begin{equation}
Q=\int \left( \partial _{R}Q\right) dR+\left( \partial _{\tau }Q\right)
d\tau .  \label{21}
\end{equation}

Using the identity $\partial _{\tau }Q=-\frac{4ME}{R}-\frac{\tau }{R}%
\partial _{R}Q$, once again, Eq. (\ref{21}) can be rewritten as

\begin{equation}
Q=\frac{1}{2}\int \frac{\partial _{R}Q}{R}d(R^{2}-\tau ^{2})-\frac{4ME}{R}%
\int d\tau .  \label{22}
\end{equation}

It is obvious that the second term is real in Eq. (\ref{22}). However, after
inserting Eq. (\ref{20}) into Eq. (\ref{22}), we see that the imaginary
contribution to the action comes only from the first term since it has pole
at the horizon. Thus, the complex path integration method \cite{angh1,pad2}
for the pole located at the horizon ($R=\tau $) yields

\begin{equation}
ImQ_{-}|_{horizon}=0,  \label{23}
\end{equation}

\begin{align}
ImQ_{+}|_{horizon}& =4\pi ME.  \label{24}
\end{align}

Therefore, the probabilities of the ingoing/outgoing massive vector
particles become

\begin{equation}
\Gamma _{absorption}=e^{-\frac{2}{\hbar }ImQ_{-}|_{horizon}}=1,  \label{25}
\end{equation}
\begin{equation}
\Gamma _{emission}=-e^{-\frac{2}{\hbar }ImQ_{+}|_{horizon}}=e^{-8\pi ME}.
\label{26}
\end{equation}

It is worth noting that the above results are in full agreement with the
semiclassical QTM \cite{Vanzo1}, which expects a 100\% chance for the
ingoing particles to enter the BH, i.e., $\Gamma _{absorption}=1$, and
thereupon computes the probability of the outgoing (tunneling) particles, $%
\Gamma _{emission}$.

The tunneling rate is then computed by

\begin{equation}
\Gamma =\frac{\Gamma _{emission}}{\Gamma _{absorption}}=e^{-8\pi ME}.
\label{27}
\end{equation}

Now, recalling the Boltzmann factor (see for example \cite{Vanzo1}), $\Gamma
=e^{-\beta E}=e^{-8\pi ME}$, where $\beta $ is the inverse temperature we
can recover the original Hawking temperature of \ Schwarzchild BH: 
\begin{equation}
T\equiv T_{H}=\frac{f^{\prime }(r_{h})}{4\pi }=\frac{1}{8\pi M}.  \label{28}
\end{equation}

\section{Quantum Tunneling of Massive Vector Particles From Schwarzchild BH
in DL Coordinates}

In 1933, Georges Lemaitre \cite{dl} found a coordinate system $\left( \tilde{%
\tau},\tilde{R},\theta,\varphi\right) $ that removes the coordinate
singularity at the Schwarzchild BH is given by

\begin{equation}
ds^{2}=-d\tilde{\tau}^{2}+\frac{d\tilde{R}^{2}}{F}+4M^{2}F^{2}(d\theta
^{2}+B^{2}d\varphi ^{2}),  \label{29}
\end{equation}%
where 
\begin{equation}
F=\left[ \frac{3}{4M}(\tilde{R}-\tilde{\tau})\right] ^{\frac{2}{3}}.
\label{30}
\end{equation}

The event horizon in the DL coordinates corresponds to $F$ $=$ $1$ or $%
\tilde{R}=\frac{4M}{3}+\tilde{\tau}$. Furthermore, the Killing vector reads

\begin{equation}
\xi ^{\mu }=\left[ 1,1,0,0\right] ,  \label{31}
\end{equation}

and it leads to the following particle energy \cite{ding}: 
\begin{equation}
E=-\xi ^{\mu }\partial _{\mu }S=-\left( \partial _{\tau ^{\ast }}+\partial
_{R^{\ast }}\right) S.  \label{32}
\end{equation}

In this coordinate system, PEs (\ref{6}) with the ans\"{a}tz (\ref{12}) are
given by%
\begin{multline}
c_{0}\left( -\left( \partial _{\varphi }S_{0}\right)
^{2}-4F^{3}M^{2}B^{2}\left( \partial _{\tilde{R}}S_{0}\right)
^{2}-4m^{2}F^{2}M^{2}B^{2}-B^{2}\left( \partial _{\theta }S_{0}\right)
^{2}\right) +4c_{1}F^{3}M^{2}B^{2}\left( \partial _{\tilde{R}\tilde{\tau}%
}S_{0}\right)  \label{33} \\
+c_{2}B^{2}\left( \partial _{\tilde{\tau}\theta }S_{0}\right) +c_{3}\left(
\partial _{\tilde{\tau}\varphi }S_{0}\right) =0,
\end{multline}

\begin{multline}
4F^{2}M^{2}B^{2}c_{0}\left( \partial _{\tilde{R}\tilde{\tau}}S_{0}\right)
+c_{1}\left( 4m^{2}F^{2}M^{2}B^{2}-4F^{2}M^{2}B^{2}\left( \partial _{\tilde{%
\tau}}S_{0}\right) ^{2}+B^{2}\left( \partial _{\theta }S_{0}\right)
^{2}+\left( \partial _{\varphi }S_{0}\right) ^{2}\right)  \label{34} \\
-c_{2}B^{2}\left( \partial _{\tilde{R}\theta }S_{0}\right) -c_{3}\left(
\partial _{\tilde{R}\varphi }S_{0}\right) =0,
\end{multline}

\begin{multline}
4F^{2}M^{2}B^{2}c_{0}\left( \partial _{\tilde{\tau}\theta }S_{0}\right)
-4c_{1}F^{3}M^{2}B^{2}c_{1}\left( \partial _{\tilde{R}\theta }S_{0}\right)
\label{35} \\
+c_{2}[4m^{2}F^{2}M^{2}B^{2}+\left( \partial _{\varphi }S_{0}\right)
^{2}-4F^{2}M^{2}B^{2}\left( \partial _{\tilde{\tau}}S_{0}\right)
^{2}+4F^{3}M^{2}B^{2}\left( \partial _{R}S_{0}\right) ^{2}]-c_{3}\left(
\partial _{\varphi \theta }S_{0}\right) =0,
\end{multline}

\begin{multline}
4F^{2}M^{2}Bc_{0}\left( \partial _{\tilde{\tau}\varphi }S_{0}\right)
-4F^{3}M^{2}Bc_{1}\left( \partial _{\tilde{R}\varphi }S_{0}\right)
-c_{2}B\left( \partial _{\varphi \theta }S_{0}\right)  \label{36} \\
+\left[ 4m^{2}F^{2}M^{2}B+B\left( \partial _{\theta }S_{0}\right)
^{2}-4F^{2}M^{2}B\left( \partial _{\tilde{\tau}}S_{0}\right)
^{2}+4F^{3}M^{2}B\left( \partial _{\tilde{R}}S_{0}\right) ^{2}\right]
c_{3}=0.
\end{multline}

Now, one can read the non-zero elements of the coefficient matrix $\aleph
\left( c_{0},c_{1},c_{2},c_{3}\right) ^{T}=0$ ($\aleph $ is another $4\times
4$ matrix) as follows

\begin{align}
\aleph _{11}& =\left[ -\left( \partial _{\varphi }S_{0}\right)
^{2}-4F^{3}M^{2}B^{2}\left( \partial _{\tilde{R}}S_{0}\right)
^{2}-4m^{2}F^{2}M^{2}B^{2}-B^{2}\left( \partial _{\theta }S_{0}\right) ^{2}%
\right] ,\   \notag \\
\aleph _{12}& =4F^{3}M^{2}B^{2}\left( \partial _{\tilde{R}\tilde{\tau}%
}S_{0}\right) ,\text{ \ }\aleph _{21}=4F^{2}M^{2}B^{2}\left( \partial _{%
\tilde{R}\tilde{\tau}}S_{0}\right) ,  \notag \\
\aleph _{13}& =B^{2}\left( \partial _{\tilde{\tau}\theta }S_{0}\right) ,%
\text{ \ }\aleph _{31}=4F^{2}M^{2}B^{2}\left( \partial _{\tilde{\tau}\theta
}S_{0}\right) ,  \notag \\
\aleph _{14}& =\left( \partial _{\tilde{\tau}\varphi }S_{0}\right) ,\text{ \ 
}\aleph _{41}=4F^{2}M^{2}B\left( \partial _{\tilde{\tau}\varphi
}S_{0}\right) ,  \notag \\
\aleph _{22}& =\left[ 4m^{2}F^{2}M^{2}B^{2}-4F^{2}M^{2}B^{2}\left( \partial
_{\tilde{\tau}}S_{0}\right) ^{2}+B^{2}\left( \partial _{\theta }S_{0}\right)
^{2}+\left( \partial _{\varphi }S_{0}\right) ^{2}\right] ,  \notag \\
\aleph _{23}& =-B^{2}\left( \partial _{\tilde{R}\theta }S_{0}\right) ,\text{
\ }\aleph _{32}=-4F^{3}M^{2}B^{2}\left( \partial _{\tilde{R}\theta
}S_{0}\right) ,  \notag \\
\aleph _{24}& =-\left( \partial _{\tilde{R}\varphi }S_{0}\right) ,\text{ \ }%
\aleph _{42}=-4F^{3}M^{2}B\left( \partial _{\tilde{R}\varphi }S_{0}\right) ,
\notag \\
\aleph _{33}& =[4m^{2}F^{2}M^{2}B^{2}+\left( \partial _{\varphi
}S_{0}\right) ^{2}-4F^{2}M^{2}B^{2}\left( \partial _{\tilde{\tau}%
}S_{0}\right) ^{2}+4F^{3}M^{2}B^{2}\left( \partial _{R}S_{0}\right) ^{2}], 
\notag \\
\aleph _{34}& =-\left( \partial _{\varphi \theta }S_{0}\right) ,\text{ \ }%
\aleph _{43}=-B\left( \partial _{\varphi \theta }S_{0}\right) ,  \notag \\
\aleph _{44}& =\left[ 4m^{2}F^{2}M^{2}B+B\left( \partial _{\theta
}S_{0}\right) ^{2}-4F^{2}M^{2}B\left( \partial _{\tilde{\tau}}S_{0}\right)
^{2}+4F^{3}M^{2}B\left( \partial _{\tilde{R}}S_{0}\right) ^{2}\right] .
\label{37}
\end{align}

Inserting the following ans\"{a}tz for $S_{0}$ 
\begin{equation}
S_{0}=\widetilde{Q}(\tilde{\tau},\tilde{R})+k(\theta )+j\varphi ,  \label{38}
\end{equation}

into Eq. (\ref{37}), and subsequently using the energy condition (\ref{32}),
namely:

\begin{equation}
\partial _{\tilde{\tau}}\widetilde{Q}=-(E+\partial _{\tilde{R}}\widetilde{Q}%
),  \label{39}
\end{equation}

we get solutions for $\partial_{\tilde{R}}\widetilde{Q}$ from $det\aleph=0$:

\begin{equation}
det\aleph =-\frac{M^{2}F^{2}Bm^{2}}{1024}\left\{ B^{2}\left( \partial
_{\theta }k\right) ^{2}+j^{2}+4M^{2}F^{2}B^{2}\left[ m^{2}+F\left( \partial
_{\tilde{R}}\widetilde{Q}\right) ^{2}-\left( E+\partial _{\tilde{R}}%
\widetilde{Q}\right) ^{2}\right] \right\} ^{3}=0,  \label{40}
\end{equation}

as follows 
\begin{equation}
\partial _{\tilde{R}}\widetilde{Q}_{\pm }=\frac{EMBF\pm \sqrt{%
E^{2}M^{2}F^{3}B^{2}-\left( F-1\right) \left[ m^{2}F^{2}M^{2}B^{2}+\frac{1}{4%
}B^{2}\left( \partial _{\theta }k\right) ^{2}+\frac{j^{2}}{4}\right] }}{%
\left( F-1\right) FBM}.  \label{41}
\end{equation}

Using the energy expression (\ref{39}) in the definite integration of $%
\widetilde{Q}:$ 
\begin{equation}
\widetilde{Q}=\int \partial _{_{\tilde{R}}}\widetilde{Q}d\tilde{R}+\partial
_{\tilde{\tau}}\widetilde{Q}d\tilde{\tau},  \label{42}
\end{equation}

we obtain 
\begin{align}
\widetilde{Q}& =\int \partial _{\tilde{R}}\widetilde{Q}d\left( \tilde{R}-%
\tilde{\tau}\right) -E\int d\tilde{\tau},  \notag \\
& =2M\int \frac{\partial _{\tilde{R}}\widetilde{Q}}{\sqrt{F}}dF-E\int d%
\tilde{\tau},  \label{43}
\end{align}

where $dF=\frac{1}{2M}\sqrt{F}d(\tilde{R}-\tilde{\tau})$ [see Eq. (\ref{30}%
)]. It is clear that the second integral of (\ref{43}) results in real
values, which means that it does not give any contribution to the imaginary
part of the action. However, after substituting Eq. (\ref{41}) into Eq. (\ref%
{43}), one can see that the first integral of Eq. (\ref{43}) has a pole at
the horizon ($F=1$), and evaluating it around the pole yields 
\begin{equation}
ImQ_{+}|_{horizon}=4\pi ME,  \label{44}
\end{equation}

and trivially

\begin{equation}
ImQ_{-}|_{horizon}=0.  \label{49}
\end{equation}%
The above results are fully consistent with Eqs. (\ref{23}) and (\ref{24}).
Consequently, they admit the same tunneling rate given in Eq. (\ref{27}). In
short, using the quantum tunneling of the massive vector particles in the DL
coordinate system, we have managed to rederive the Hawking temperature ($%
T_{H}=\frac{1}{8\pi M}$) of the Schwarzchild BH.

\section{ Conclusion}

In this paper, we have used the PE (\ref{6}) in order to compute the HR of
the massive vector particles tunneling from the Schwarzchild BH given in two
different (KS and DL) regular dynamic coordinate systems. In addition to the
HJ and the WKB approximation methods, particle energy definitions played
crucial role in our computations. The original Hawking temperature of the
Schwarzschild BH is impeccably obtained in the both coordinate systems.
Thus, we have shown that HR is independent of the selected coordinate
system. The latter remark was also highlighted in \cite{CQG}.

Finally, we plan to extend our present study to the HR of the massive spin-1
particles that experience the minimum length effect \cite%
{mle1,mle2,mle3,mle4}, which is governed by the GUP (generalized uncertainty
principle) \cite{gup1,gup2,gup3,gup4,gup5,gup6,gup7}. Such a study requires
the following modification in the PE \cite{gup8}:

\begin{equation}
\frac{1}{\sqrt{-g}}\partial_{\mu}\left[\left( 1-\frac{l_{p}^{2}}{3}%
\partial_{\mu }^{2}\right) \sqrt{-g}\Phi^{\nu\mu}\right]+\frac{m^{2}}{%
\hbar^{2}}\Phi^{\nu }=0,  \label{50}
\end{equation}

where

\begin{equation}
\Phi_{\nu\mu}=\partial_{\nu}\Phi_{\mu}-\partial_{\mu}\Phi_{\nu}-\partial_{%
\nu }\frac{l_{p}^{2}}{3}\partial_{\mu}^{2}\Phi_{\mu}+\partial_{\mu}\frac{%
l_{p}^{2}}{3}\partial_{\nu}^{2}\Phi_{\nu},  \label{51}
\end{equation}
in which $l_{p}$ denotes the Planck length. This problem may reveal more
information compared to the present case. This is going to be our next study
in the near future.

\section*{Acknowledgement}

The authors are grateful to the editor and anonymous referees for their
valuable comments and suggestions to improve the paper.

\end{document}